\documentstyle[prl,aps,psfig,multicol]{revtex}
\draft
\newcommand{\be}{\begin{equation}}
\newcommand{\ee}{\end{equation}}

\newcommand{\bea}{\begin{eqnarray}}
\newcommand{\eea}{\end{eqnarray}}
\begin{document}
\title{Bounds on tau neutrino magnetic moment and charge radius from
Super-K and SNO  observations }
\author{
Anjan S. Joshipura and Subhendra Mohanty}
\address{ Physical Research Laboratory,
Navrangpura, Ahmedabad - 380 009, India}

\maketitle

\begin{abstract}
Neutrinos can scatter electrons in water detectors through their
magnetic moments and charge radii in addition to the charged and
neutral currents channels. The recent solar neutrino charged
current event rates announced by SNO with the earlier solar and
atmospheric neutrino observations from Super-Kamiokande allows us
to put upper bounds of $\mu < 10^{-9} \mu_B$ on neutrino magnetic
moments and $<r^2> < 10^{-31} cm^2$ on the neutrino charge radii .
For the electron and muon neutrinos these bounds are comparable
with existing bounds but for tau neutrinos  these bounds are three
orders of magnitude more stringent than earlier terrestrial bounds
.  These bounds are independent of any specific model of neutrino
oscillations.
\end{abstract}

\begin{multicols}{2}

Among the electro-magnetic form factors of three  neutrinos
flavors,  the magnetic moment and charge radius of the $\nu_\tau$
are the least constrained in terrestrial experiments. This is
because of the non-availability of a copious flux of $\nu_\tau$'s
in terrestrial sources like reactors and colliders. In contrast to
the terrestrial sources, the Sun provides us with an intense beam
of tau neutrinos which result from the conversion of the original
electron neutrinos. Combination of several experimental
observations can be used to infer the existence of $\nu_\tau$ from
the Sun and quantify its flux: (a) The recent observations of
$\nu_e$'s at the Sudbury Neutrino Observatory (SNO)\cite{sno}
establishes that the $^8B$ neutrinos are reduced to a fraction
$P_{ee}=0.347$ due to either vacuum oscillations or MSW conversion
of solar neutrinos. The neutrino flux observed at SNO also implies
that the earlier Super-Kamioka \cite{sk1} flux measurements from
the $\nu_e$ elastic scattering contained contribution from the
neutral current also. This contribution is consistent with what is
expected from the $\nu_e$ conversion to active flavours
$\nu_{\mu,\tau}$. (b) Super-Kamiokande \cite{sk2} observations of
atmospheric neutrinos show that $\nu_\mu$ oscillates with
oscillation length $\sim 13000 km$ (the earth diameter) and with
maximal mixing ($Sin^2 2 \theta_{23} \sim 1/2$). The absence of
the matter effects in this oscillations \cite{sk3} favour the
conversion of $\nu_\mu$ to $\nu_{\tau}$. The presence of the large
amplitude $\nu_\mu-\nu_{\tau}$ oscillations imply that significant
fraction of the solar neutrinos convert to $\nu_{\tau}$. Roughly,
half of the converted fraction $(1-P_{ee})$ of $\nu_e$'s
contribute to $\nu_\mu$' and the remaining half to $\nu_\tau$'
flux. Thus the neutrino beam from the sun (in the  $5-15 MeV$
energy range) appears as composed of the mixture of
$\nu_e,\nu_\mu,\nu_\tau$ in approximately the ratio $0.347 : 0.326
: 0.326$ respectively (the ratios being slightly different for
lower energy neutrinos) at the earth. The one-third fraction of
$\nu_\tau$'s of the solar neutrino flux can therefore be utilized
to study the electro-magnetic form factors of tau neutrinos.

The Super-Kamiokande experiment observes the elastic scattering
(ES) of electrons in the water target which can be caused by both
charged current (CC) and neutral current (NC) processes ($\nu_x +
e^- \rightarrow \nu_x + e^-$ where $\nu_x= \nu_e, \nu_\mu $ or
$\nu_\tau$). The SNO experiment uses a heavy water target and can
differentiate the charged current reaction( $\nu_e + d \rightarrow
p+p+e^-$)  from the neutral current  ( $\nu_x + d \rightarrow
p+n+\nu_x$) and the electron scattering reactions. One can
subtract the charged current rates of of SNO from the total
elastic scattering rates of Super-K \cite{sk2} to put bounds on
scattering processes other than the charged and neutral current
weak interactions.  In this paper, we study the elastic scattering
of electrons at SK due to the magnetic moments and charge radii
of the solar neutrinos. Using the recent SNO results \cite{sno}
along with the elastic scattering rates from Super-K \cite{sk2} we
obtain the following upper bounds (at $90\% C.L$) on neutrino
electro-magnetic form factors (in the weak interaction basis):

\noindent {\it Dirac moments}
 \bea
\mu_\tau,\mu_\mu &<&  6.73 ~( 5.77)~ \times 10^{-10} \mu_B,
\nonumber\\
 \mu_e &<&6.45 ~( 5.65)~ \times 10^{-10} \mu_B ,
 \label{1dirac}
 \eea
{\it Transition moments}
 \bea
 \mu_{e \tau}, \mu_{e \mu} &<& 4.66 ~( 4.04)~ \times 10^{-10} \mu_B,
 \nonumber\\
 \mu_{\mu \tau} &<& 4.76 ~ ( 4.08)~ \times 10^{-10} \mu_B,
 \label{transition}
 \eea
 {\it Charge radii}
 \bea
 |\langle r^2 \rangle_{\nu_\tau}|,|\langle r^2 \rangle_{\nu_\mu}| &<&
 2.08
  ~(1.53)~ \times 10^{-31} cm^2, \nonumber\\
|\langle r^2 \rangle_{\nu_e}| &<& 6.86
 ~(5.26)~ \times 10^{-32} cm^2.
\eea
 where the main contribution to the errors is the theoretical uncertainty in
 the $^8B$ neutrino flux in the standard solar model
 \cite{bahcall}. The numbers displayed in the bounds (1-3) are
 calculated assuming an uncertainty of $\pm 20\%$ in the SSM
 neutrino flux. In the brackets we give the numbers assuming a SSM
 flux uncertainty of $5\%$ .

 These bounds are independent of whether the conversion
mechanism of $\nu_e$ from the Sun is vacuum oscillations, or either large
or small angle MSW \cite{fits} but they assume three light neutrinos.
Bounds similar to above can be obtained even if $\nu_e$ mixing with
sterile state consistent with present experiments \cite{barger} is
allowed.
Likewise, one obtains bounds of similar magnitudes if the solar
neutrinos convert to anti- neutrinos through spin-flip in solar magnetic field
\cite{rsfp}.

 Our bound on the tau neutrino Dirac magnetic moment is
about three orders of magnitude more stringent than the previous
best bound $ \mu_\tau < 5.4 \times 10^{-7} \mu_B (90\%
C.L)$\cite{cooper} obtained by a terrestrial experiment (from
$\nu_\tau e  $ scattering by tau neutrinos obtained from $D_s$
decay) . The bound on $\mu_{\nu_\mu} $ is comparable with the
bound $\mu_{\nu_\mu} < 7.4 \times 10^{-10} \mu_B$ obtained from $\nu_\mu
e$ elastic scattering measurements at LAMF \cite{lamf}. Our bound
on $\mu_{\nu_e}$ is weaker by a factor of five compared to the
earlier bound $\mu_{\nu_e} < 1.5 \times 10^{-10} \mu_B$ obtained
by analysis of the spectral distortion of electron scattering by
solar neutrinos at Super-K \cite{vogel1}.

The charge radius of $\nu_e $ has been bounded by the LAMF
experiment ( $<r^2> =(0.9 \pm 2.7 ) 10^{-32} cm^2)$ and for the
muon neutrino the bound from scattering experiment \cite{vilain}
is ($<r^2>_\mu < 0.6 \times 10^{-32} cm^2$). To our knowledge there are
no bounds on the charge radius of $\tau $ neutrinos from
scattering experiments.

 Our method is similar to that of \cite{pulido} who
also used the total Super-K scattering rates to put bounds on
neutrino electro-magnetic form factors.  Our analysis differs from
\cite{vogel1} and \cite{pulido} in the following ways:(a) we have
included the recent SNO CC rates to subtract out the weak
interaction part from the Super-K elastic scattering rate. This
avoids assumption \cite{pulido} about the relative strength of the
charged and neutral current contribution at Super-K (b) our
analysis is more conservative as we have also included a possible
$20\%$ error in the theoretical  $^8 B$ flux prediction
\cite{bahcall} which as emphasized in \cite{barger} is the main
source of uncertainty in extracting bounds from Super-K and SNO
observations , and mainly (c) we have used the results from
atmospheric neutrinos to establish that $\sim 1/3$ of the solar
neutrino flux consists of $\nu_\tau$ which enables us to put much
more stringent bounds on the $\nu_\tau$ magnetic moment and charge
radius than was possible in earlier terrestrial experiments.

From the analysis of atmospheric neutrinos \cite{sk1} and the
$\nu_e$ disappearance experiment at CHOOZ \cite{chooz} we can
write the mixing matrix $U$ for the three neutrino generations as
 \bea
\pmatrix{
  \nu_e \cr
  \nu_\mu\cr
  \nu_\tau}
= \pmatrix{
  c_1& s_1 & 0 \cr
  -c_2 s_1 & c_2 c_1 & s_2\cr
  -s_2 c_1& s_2 c_1& c_2}
 ~~\pmatrix{
  \nu_1 \cr
  \nu_2\cr
  \nu_3}
 \label{u}
\eea

 Here $(c_1,s_1)$ denote the mixing relevant for
the solar neutrinos and $(c_2,s_2)$ are
the mixings relevant for the atmospheric neutrinos.
 The $U_{e3}$ element is set to zero from the  CHOOZ
observation that disappearance of $\nu_e$ is less than $4\%$ (with
sensitivity to $\delta m^2 \sim 10^{-3} eV^2$ in the atmospheric
range), which implies that $U_{e3}^2 (1-U_{e3}^2) < 0.01$. Using
this constraint with the atmospheric neutrino observation of
maximal mixing between $\nu_\mu$ and $\nu_\tau$ ( $U_{\mu 3}^2 =
U_{\tau 3}^2 =0.5$)  then (by unitarity)  implies that  $U_{e 3}^2
< 0.01$. We have set the $U_{e3}$ entry in (\ref{u}) to be zero
for simplicity ( the $1\%$ error incurred here is negligible
compared the experimental errors ($\sim 5\%$) and theoretical flux
uncertainty ($20 \%$)).

 The conversion probability to $\nu_\tau$ and $\nu_\mu$ of $\nu_e$ produced in the solar
 core can be derived using (\ref{u}) and turns out to be of the form
 \bea
 P_{e \mu} &=& c_2^2~ (1-P_{ee}),\nonumber\\
  P_{e \tau} &=& s_2^2 ~(1-P_{ee})
 \label{probs}
 \eea
 The assumptions made in deriving
 (\ref{probs}) are  (a) in case of vacuum oscillations , the
 oscillation length scale $l_{23}$ associated with the $(\nu_2,\nu_3)$ mass difference
 ( $\delta m^2_{23})$ is
 smaller than the earth sun distance or (b) in case of matter
 induced conversion of $\nu_e$ , the matter potential in the sun
 $\sqrt{2} G_F N_e << \delta m_{23}^2 /2 E $ and the oscillation
 length $l_{23} << R_\odot.$  These conditions are met by the
 atmospheric neutrino oscillation length scale $l_{23}\sim 2 R_\oplus = 13,000
 km $ with the associated mass scales $\delta m_{23}^2
 \sim 10^{-3} eV^2$ \cite{sk1}.
The form of eq.(\ref{probs}) is independent  of the mechanism of
the solar $\nu$ conversion but the expression for  $P_{ee}$ depends on the
specific mechanism which could  either be vacuum oscillations or
large or small angle MSW depending on the $(\nu_1,\nu_2)$ mass
difference. Simultaneous use of the SNO and SK results allows
direct extraction of $P_{ee}$ from experiments and makes the
following analysis independent of whether the specific mechanism
of conversion MSW or VO. The analysis differs if there is conversion to
anti-neutrinos by some resonant helicity flip mechanism or if
there is a significant conversion to sterile neutrinos. These two
possibilities are discussed separately in the final sections.

Supposing the $\nu_e$ produced in the Sun are converted to
$\nu_\mu$'s and $\nu_\tau$' with probabilities given by
(\ref{probs}) ,  the rates of elastic scattering events at Super-K
and charged current events at SNO can be written as
  \bea
{\phi^{ES}_{SK}\over \phi_{SSM}} &=& \langle P_{ee} \rangle +
\langle {\sigma_{\nu_\mu}\over \langle \sigma_{\nu_e} \rangle}
(1-P_{ee})\rangle + \langle {\sigma^\gamma_{\nu_e}
\over \langle \sigma_{\nu_e} \rangle} P_{ee} \rangle \nonumber\\&+&
\langle
{\sigma^\gamma_{\nu_\mu} \over \langle \sigma_{\nu_e} \rangle}
c_2^2(1- P_{ee})+ \langle {\sigma^\gamma_{\nu_\tau }
\over \langle \sigma_{\nu_e} \rangle} s_2^2(1- P_{ee}) \rangle
\label{rsk} \eea and
\be
{\phi^{CC}_{SNO}\over \phi_{SSM}} = \langle P_{ee} \rangle
\label{rsno}~, \ee
 where $\sigma_{\nu_e}$ ($\sigma_{\nu_\mu}$) denote the weak
 $(\nu_e\;e)$ ($(\nu_\mu\; e)$ and $(\nu_\tau\; e)$) elastic
scattering cross sections.
 The photon mediated scattering by a neutrino $\nu_\alpha$ ,
 is denoted by $\sigma^\gamma_{\nu_\alpha}$ (where  $\alpha=e,\mu,\tau$).
These
 cross sections are proportional to the  diagonal  or transition
  magnetic moment
 $\mu_{\alpha \beta}$ or the charge radii $<r^2>_\alpha$ of
 $\nu_\alpha$.

The cross section for the elastic scattering $\nu_\alpha + e^-
\rightarrow \nu_\beta +e^-$ due to neutrino magnetic moment is
given by the expression \cite{vogel2}
\be
{d\sigma_{\nu_\alpha}^\gamma \over dT} = \mu_{\alpha \beta}^2~ (
\alpha_{em}) ( {1\over T} -{1\over E_\nu})
 \label{mm}
 \ee
where $T$ is the recoil energy of the scattered electron. The
magnetic moments are defined in the flavour basis corresponding to
the diagonal charged currents and differ from the corresponding
moment in the neutrino mass basis \cite{vogel1}.
 The conical brackets in (\ref{rsk}-\ref{rsno}) denote
averaging over the neutrino energy spectrum from the sun and the
detector response function of the detector. We make use of the
observation \cite{sno} and \cite{sk2} that  $P_{ee}$ is energy independent
in the neutrino  energy range probed in these experiments. This allows
the simplification  $\langle \sigma P_{ee}
\rangle \simeq \langle \sigma \rangle \langle P_{ee}\rangle $ in (\ref{rsk}).
 We
average the cross sections for the weak interaction scatterings
and the magnetic moment (and charge radius) mediated scatterings
by using the analytical form for the
 $^8B$ spectrum of solar neutrinos given in \cite{bh}. We
incorporate the  error involved in the measurement of the recoil
 electron  energy $T$ by averaging
the expressions over a detector response function $r(T,T^\prime)$
given in \cite{lisi}. We chose the threshold energy  to be $E_\nu
\geq 8.5 MeV$ since as shown by \cite{lisi} with this threshold
the rates at Super-K and SNO can be compared (although the two
experiments have different thresholds and different detector
responses) and we take the upper limit for $E_\nu =15 MeV$. The
averaged  magnetic moment scattering cross section (\ref{mm})
turns out to be
\be
\langle \sigma^\gamma(\nu_\alpha e^- \rightarrow \nu_\beta e^-)
\rangle= \kappa_{\alpha \beta}^2~ 4.877 \times 10^{-27} cm^2
\label{sigmamu}
 \ee
 where $\kappa_{\alpha \beta}$ are the diagonal or transition
 magnetic moments in units of the Bohr magnetron .
 The
average values of the $\nu_e$ and $\nu_\mu$ weak elastic
scattering cross sections (\ref{sigma}) with $e^-$  turn out to be
 \bea \langle \sigma_{\nu_e}
\rangle &=& 9.108 \times 10^{-45} cm^2 ,\nonumber\\ \langle
\sigma_{\nu_\mu} \rangle &=& 1.343 \times 10^{-45} cm^2.
 \label{sigma}
\eea
 Substituting (\ref{sigmamu}) and (\ref{sigma})
in (\ref{rsk}) and (\ref{rsno}) and using the experimental results
for the elastic scattering rates from Super-K \cite{sk2}
\be
\phi^{ES}_{SK} = (0.451 \pm 0.017) \times \phi_{SSM} \label{expsk}
\ee and the charged current rates from SNO\cite{sno}
\be
\phi^{CC}_{SNO} = (0.347 \pm 0.029) \times \phi_{SSM} \ee
along with the theoretical solar neutrino flux prediction \cite{bahcall}
 \be
\phi_{SSM} = (5.05 \times 10^{-6} cm^{-2} s^{-1}) + 20\% -16\%
\label{ssm}
\ee
and with the experimental determination of the $(\nu_2, \nu_3)$ mixing
angle
from atmospheric neutrinos \cite{sk1}
 \be
 Sin^2 \theta_2 = 0.5 \pm 0.05
 \label{theta}
 \ee
 we obtain the following limits on neutrino magnetic moments (with
 the errors shown at one $ \sigma$).

\noindent {\it Dirac moments}
 \bea
 \mu_\tau^2 =\mu_\mu^2&=&(4.42 \pm 24.92 )\times 10^{-20} \mu_B^2, \nonumber\\
 \mu_{e}^2&=&(4.16 \pm 22.84 )\times 10^{-20} \mu_B^2,
 \label{dirac}
 \eea
 {\it Transition moments}
 \bea
 \mu_{e \tau}^2=\mu_{e \mu}^2 &=& (2.14 \pm 11.92) \times 10^{-20}
 \mu_B^2 ,\nonumber\\
 \mu_{\nu_{\mu \tau}}^2 &=&(2.11 \pm 12.46 )\times 10^{-20} \mu_B^2.
 \label{trans}
 \eea
  The transition moments have better bounds as they contribute to
 the elastic scattering (\ref{rsk}) through two channels whereas
 the Dirac moments contribute to one term at a time in (\ref{rsk}).
 The bounds (\ref{dirac}) and (\ref{trans}) translate to the upper
 bounds (1-2) at $90\% C.L$ .
 The largest source of error in obtaining these bounds is the
 large uncertainty $-16\%$ to $+20 \%$
 in the $^8 B$ flux prediction of the standard
 solar model \cite{bahcall}. If the $^8 B$ flux can be determined
 from solar neutrino experiments themselves to an accuracy of $\pm
 5 \%$ then the bounds on the magnetic moments improve as
 shown by bracketed numbers in equations (1-3).

 Using a similar analysis one can establish bounds on the charge
 radius scattering by neutrinos.
 This scattering unlike the neutrino magnetic moment one preserves
 the helicity of the neutrinos and it can therefore interfere with
 the weak interaction elastic scattering amplitude. The
 contribution of this interference term to the total elastic
 scattering cross section
 for the
 $(\nu_\alpha ~e)$  scattering is given by \cite{vogel2}
 \bea
 {d \sigma^\gamma_{\nu_\alpha} \over dT} &=& \langle r^2 \rangle_\alpha~
{\sqrt{2}\over 3}
 G_F \alpha_{em} m_e ( (g_V+g_A) \nonumber\\
&+& (g_V-g_A) (1-{T \over E_\nu})^2 - g_V {m_e T \over E_\nu^2} )
 \eea

Following the same averaging procedure as for the magnetic moment
case we find the expression for the flux and detector response
averaged cross sections due to charge radius scattering are
 \bea
\langle \sigma^\gamma_{\nu_e} \rangle &=& 2.96 \times 10^{-14}
\langle r^2 \rangle_{\nu_e} ,\nonumber\\ \langle
\sigma^\gamma_{\nu_{\mu},\nu_{\tau}} \rangle &=& -1.06 \times
10^{-14} \langle r^2 \rangle_{\nu_\mu,\nu_\tau}.
 \label{sigmacr}
\eea
Using the averaged cross sections (\ref{sigmacr}) in equation
(\ref{rsk}) and using the experimental rates  (\ref{expsk}-\ref{theta})
and theoretical
prediction (\ref{ssm}) we find that the limits on
neutrino charge radii are
 \bea
 \langle r^2 \rangle _{\nu_\mu}=\langle r^2 \rangle _{\nu_\tau}&=&
 (-2.03 \pm 11.46 )\times 10^{-32} cm^2,\nonumber\\
\langle r^2 \rangle _{\nu_e}&=&
 (0.686 \pm 3.77)\times 10^{-32} cm^2.
\eea These bounds can be converted to  upper bounds on the
neutrino charge radii as given in (3) .

{\bf Bounds in case of conversion to sterile neutrinos}: The
analysis so far assumed only three light neutrinos. While this
possibility seems most favoured, significant admixture of sterile
state is still allowed \cite{barger}. Suppose only a fraction
$Sin^2 \alpha$ of the converted solar neutrinos are active
($\nu_\mu$ or $\nu_\tau$). In that case the expression (\ref{rsk})
must be modified by replacing the factor $(1-P_{ee})$ by $Sin^2
\alpha (1-P_{ee})$ . As shown in \cite{barger} one can combine the
average rates from the Cl and Ga experiments with the SK and SNO
results to limit put limits on $Sin^2 \alpha$. The SSM flux
prediction along with the central values of experimental rates
implies \cite{barger} no sterile mixing, i.e. $Sin^2\alpha=1$.
However  at $1 \sigma$, a $Sin^2\alpha$ as small as 0.3 is allowed
in case of the SSM. It is still possible to obtain significant
bound on magnetic moments in this case. We illustrate this taking
a specific case of $Sin^2\alpha=0.3 $ and  $20\%$ uncertainty in
the SSM flux. The bounds on the neutrino EM form factors turn out
in this case to be (with $90\%C.L$)

\noindent{\it Dirac moments}
 \bea
 \mu_\tau ,~\mu_\mu&<& 1.68 \times 10^{-9} \mu_B ,\nonumber\\
 \mu_{e}&<& 0.865 \times 10^{-9} \mu_B,
 \label{dirac2}
 \eea
 {\it Transition moments}
 \bea
 \mu_{e \tau},~\mu_{e \mu} &<&7.63 \times 10^{-10}
 \mu_B, \nonumber\\
 \mu_{\nu_{\mu \tau}} &<& 1.18\times 10^{-9} \mu_B,
 \label{trans2}
 \eea
 {\it Charge radii }
\bea
 |\langle r^2 \rangle _{\nu_\mu}|,~|\langle r^2 \rangle _{\nu_\tau}|&<&
  1.29 \times 10^{-30} cm^2,\nonumber\\
|\langle r^2 \rangle _{\nu_e}|&<&  1.23 \times 10^{-31} cm^2. \eea

{\bf Bounds for RSFP solution}:
 In case the correct solution of the solar neutrino problem is
resonant spin flip in the solar magnetic field then also the same
procedure we have followed can be applied. If $\nu_e$
convert to $\bar \nu_{\mu}$ due to resonant spin-flip (RSFP)
\cite{rsfp} in the sun then again due to $\bar \nu_\mu ,\bar
\nu_\tau$ mixing we will have neutrinos from the sun in the ratio
$\nu_e:\bar \nu_\mu:\bar\nu_\tau :: 0.347:0.326:0.326$. The
$\nu_\mu ,e$ weak interaction cross section in (\ref{rsk}) will be
replaced by the $\bar \nu_\mu ,e $ cross section whose value
(after averaging over the $^8B$ spectrum (\cite{bh}) and the
detector response function \cite{lisi}) turns out to be
 \bea
\langle\sigma_{\bar \nu_\mu }\rangle &=& 1.014 \times 10^{-45}
cm^2
 \label{sigmabar}
 \eea
and the charged radii cross sections for $\nu_\mu$ and $\nu_\tau$
in (\ref{sigmacr}) have to be replaced by their corresponding
anti-particle cross sections,
 \bea
 \langle
\sigma^\gamma_{\bar \nu_{\mu},\bar \nu_{\tau}} \rangle = 0.895 \times
10^{-14} \langle r^2 \rangle_{\bar \nu_\mu,\bar \nu_\tau}.
 \label{sigmacrb}
\eea
 Making these changes in (\ref{rsk}) and following the same
 procedure as above we obtain the following bounds on the
 E-M form factors of the neutrinos ($90\% C.L$):

\noindent{\it Dirac moments}
 \bea
 \mu_\tau ,~\mu_\mu&<& 0.767 \times 10^{-9} \mu_B, \nonumber\\
 \mu_{e}&<& 0.721 \times 10^{-9} \mu_B.
 \label{dirac1}
 \eea
 {\it Transition moments}
 \bea
 \mu_{e \tau},~\mu_{e \mu} &<& 0.524 \times 10^{-9}
 \mu_B ,\nonumber\\
 \mu_{\nu_{\mu \tau}} &<&0.542\times 10^{-9} \mu_B,
 \label{trans1}
 \eea
 {\it Charge radii }
\bea
 |\langle r^2 \rangle _{\bar \nu_\mu}|,
~|\langle r^2 \rangle _{\bar \nu_\tau}|&<&
3.21\times 10^{-31} cm^2,\nonumber\\ |\langle r^2 \rangle
_{\nu_e}| &<& 0.858 \times 10^{-31} cm^2. \eea
Although the RSFP solution is sensitive to the values of transition magnetic
moments and solar magnetic field, the above analysis did not make use of
the dynamics
of RSFP and only assumed that $P_{ee}$ is reduced to experimental
value
through the conversion of  $\nu_e$ to $\bar \nu_{\mu}$ or $\bar
\nu_{\tau}$ inside
the Sun.

{\bf Conclusions }: Measurements of  electro-magnetic form factors
of tau neutrinos in terrestrial experiments is limited by the
absence of a calibrated copious $\nu_\tau$ source.  We have shown
that the solar neutrino beam with roughly equal mixture of all the
neutrino species can be used for putting bounds on $\tau$
neutrinos which are far more stringent than bounds from other
terrestrial experiments. Bounds on neutrino properties can be put
from cooling rates of supernovae and helium stars \cite{raffelt}
are stronger though less reliable than scattering experiments from
calibrated sources.

\end{multicols}
\end{document}